\documentstyle[preprint,aps,eqsecnum,epsf]{revtex}
\def\Z0{${\em Z^0\/}$}

\def\r#1 {$^{#1}$}

\newcommand{\bbbar}{b\bar{b}}
\newcommand{\ccbar}{c\bar{c}}

\def\gepsfcentered#1{
  \def\testit{#1}
  \def\lbracket{[}
  \ifx\testit\lbracket
    \let\dofilecmd=\gepsfwithopt
  \else
    \let\dofilecmd=\gepsfnoopt
  \fi
  \dofilecmd}

\def\gepsfnoopt#1{
  \begin{center}
  \leavevmode
  \epsffile{#1}
  \end{center}}

\def\gepsfwithopt#1 #2 #3 #4]#5{
  \begin{center}
  \leavevmode
  \gepsfmaxx=0.94\textwidth
  \epsffile[#1 #2 #3 #4]{#5}
  \end{center}}

\newdimen\gepsfmaxx
\def\epsfsize#1#2{
  \ifnum #1 > \gepsfmaxx
    \gepsfmaxx
  \else
    #1
  \fi
}

\def \Rb  {$R_b~$}
\def \Rc  {$R_c~$}

\def \eb {$\epsilon_b~$}
\def \ebi {\epsilon_b}
\def \ec {$\epsilon_c~$}
\def \eci {\epsilon_c}

\def \euds {$\epsilon_{uds}~$}
\def \eudsi {\epsilon_{uds}}
\begin{document}
\begin{flushright}
FERMILAB-PUB-96/26-T\\
\today \\
\end{flushright}

\begin{Large}
\begin{center}

{\bf {Comments on Recent Measurements of $\rm{\bf{R_c}~and~\bf{R_b}}$ }}


\end{center}
\end{Large}
\begin{center}
\begin{large}
\bf{Isard Dunietz}\footnote[1]{Theoretical Physics Division},
\bf{Joseph Incandela}\footnote[2]{CDF experiment} \\
\end{large}
\it{Fermi National Accelerator Laboratory \\
Batavia, IL 60510, USA} \\
\begin{large}
\bf{Frederick D. Snider} \\
\end{large}
\it{The Johns Hopkins University \\
Baltimore, Maryland 21218, USA}\\
\begin{large}
\bf{Kyuzo Tesima}\footnote[3]{Department of Physics, School of Medicine} \\
\end{large}
\it{Fujita Health University \\
Toyoake, Aichi 470-11, Japan} \\
\begin{large}
\bf{Isamu Watanabe} \\
\end{large}
\it{Akita Junior College \\
46-1 Morisawa, Sakura, Shimokitate \\
Akita 010, JAPAN} \\
\end{center}

\newpage
\begin{abstract}
Discrepancies between
Standard Model predictions and experimental measurements of 
the fractions $R_c$ and $R_b$ of hadronic Z decays to charm and bottom
are investigated. 
We show that there exists a discrepancy in two complementary determinations
of $B(\overline{B}\rightarrow DX)$.
Reducing the branching ratio $B(D^0 \rightarrow K^- \pi^+)$  by $\sim15\%$ 
from currently accepted values to $(3.50\pm 0.21)\%~$ removes the
discrepancy. Since  $B(D^0 \rightarrow K^- \pi^+)$ calibrates most
charmed hadron yields, the reduced value 
also eliminates the discrepancy between the
predicted and measured values of $R_c$ and mitigates a problem in 
semileptonic $B$ decays. A reduction in $B(D^0 \rightarrow K^- \pi^+ )$ would
also mean that roughly $15\%$ of all $D^0$ and $D^+$ decays have not
been properly taken into account. It is shown that if the
missing decay modes involve multiple charged particles, they would be 
more likely to pass the requirements for lifetime $B$ tagging 
at LEP and SLC. This would mean that the charm tagging efficiency in
$Z\rightarrow \ccbar$ has been underestimated. 
As a consequence $R_b$ would need to be revised downward, potentially 
bringing it in line with the Standard Model prediction.
\end{abstract}
\newpage

\begin{large}
\section{Introduction}
\end{large}
Recent discrepancies between theoretical predictions and experimental 
measurements of the fractions $R_c$ and $R_b$ of hadronic Z decays to charm 
and bottom \cite{graziani,lepew96} could have very serious implications for
the Standard Model. It is therefore extremely important to determine whether
or not there exist possible explanations for these discrepancies which do 
not contradict the current paradigm.  
To this end we show in Section II that there now exist two complementary 
determinations of $B(\overline{B}\rightarrow DX)$\footnote{Throughout this
note, CP violation is neglected and for each process its CP-conjugate is
implied.}. 
After making reasonable adjustments to charmed baryon yields, we show that
the two estimates disagree.  We find that a $15\%$ reduction in 
$B(D^0 \rightarrow K^-\pi^+)$ to $(3.50\pm 0.21)\%~$
eliminates the problem. We then proceed to demonstrate that reducing 
$B(D^0 \rightarrow K^- \pi^+ )$ also eliminates the discrepancy between the
predicted and measured values of $R_c$ and alleviates a problem in 
semileptonic $B$ decays. One further consequence of the change in 
$B(D^0 \rightarrow K^- \pi^+ )$ would
be that roughly $15\%$ of all $D^0$ and $D^+$ decays have not
been properly taken into account.  In Section III we show that if these
missing decay modes involve multiple charged particles, they would be 
more likely to pass the requirements for lifetime $B$ tagging 
at LEP and SLC. In this case, the charm tagging efficiency in
$Z\rightarrow \ccbar$ events would have been underestimated. 
This would necessitate a downward revision in the measured value of 
$R_b$  which would bring it closer to the Standard Model prediction. The
effect could in fact be large enough to completely eliminate
the $R_b$ discrepancy.

\newpage
\begin{large}
\section{\Rc Measurements and $B(D^0 \rightarrow K^- \pi^+ )$}
\end{large}

The fraction $R_c$ of hadronic $Z$ decays to 
charm, which is predicted to be 0.172 in the 
Standard Model, has recently been measured to be $0.1598 \pm 0.0069$  
\cite{graziani,lepew96}. 
Similarly, the number of charm quarks per $B$ decay $(n_c)$
was historically 
measured to be smaller than expected~\cite{nchistory}, especially 
in view of the small measured inclusive semileptonic $B$ decay branching ratio.
Furthermore, the sum over all branching ratios of 
exclusive semileptonic $B$ decays falls significantly short of the inclusive 
$B(B\rightarrow X\ell\nu )$ measurements \cite{drellp}.

One possible explanation for these discrepancies is that a systematic
under-counting of charm has taken place. In particular, 
a common thread in these measurements is a significant reliance upon the 
measured value of $B(D^0\rightarrow K^- \pi^+)$ to calibrate charm production 
and decay for a wide range of observable decay modes. The CLEO experiment measures 
\cite{cleod0}
$$ B(D^0 \rightarrow K^- \pi^+) = (3.91 \pm 0.19) \% \; ,$$
and the 1994 Particle Data Group~\cite{pdg} cites a world average of
$$ B(D^0 \rightarrow K^- \pi^+) = (4.01 \pm 0.14) \% \; .$$
These calibrate not only the $D^0$ decay modes, but the
$D^+$ decay modes as well \cite{cleorp}, via the ratio
$$r_+ \equiv 
\frac{B(D^+\rightarrow K^-\pi^+\pi^+)}{B(D^0\rightarrow K^-\pi^+)}\;.$$
The calibration mode for $D_s$, namely $B(D_s \rightarrow \phi\pi )$, has  
also recently been tied to $B(D^0 \rightarrow K^- \pi^+ )$ in a  
model-independent fashion \cite{cleods}. 

As a result of a recent measurement by CLEO\cite{moriond} of the wrong-charm
production in flavor-tagged $B$ decays, it is now possible to determine
the right-charm branching fraction, $B(\overline{B} \rightarrow DX)$, in
two complementary ways. As one important consequence, we can
treat $B(D^0 \rightarrow K^- \pi^+ )$ as an unknown which is
determined by equating the two results for $B(\overline{B} \rightarrow DX)$. 
This exercise is carried out in the next section after we address
several concerns related to charmed baryon 
yields which result in an overall reduction in the estimate for 
weakly decaying charmed baryon production in $B$ decays.

\vskip 0.5 in
\begin{large}
\subsection{Inclusive D Yields in $\overline{B}$ Decays and
$\bf{B(D^0 \rightarrow K^- \pi^+)}$}
\end{large}

The number of charmed hadrons per $B$ decay is defined as 

\begin{equation}
n_c \equiv Y_D +Y_{D_s}+Y_{baryon_c}+2 B(\overline B\rightarrow (c\bar c)X)\;, 
\end{equation}
where the inclusive production of final states containing an arbitrary
charmed hadron $T$ is defined by

\begin{equation}
Y_T\equiv B(\overline B\rightarrow TX)+B(\overline B\rightarrow \overline TX)\;.
\end{equation}

The weakly decaying, singly charmed baryon species  
$(\Lambda_c ,\Xi^{+,0}_c , \Omega_c )$ are collectively denoted by $baryon_c$ 
while $(c\bar c)$ represents charmonia not seen as open charm.
Table I summarizes CLEO measurements with the underlying calibration 
terms factored out explicitly.  Note that the inclusive $D^+$ yield in  
$B$ decays involves $B(D^+ \rightarrow K^- \pi^+ \pi^+)$ which is  
calibrated by $D^0 \rightarrow K^- \pi^+$ \cite{cleorp} via the ratio,
\begin{equation}
r_+\equiv
\frac{B(D^+ \rightarrow K^- \pi^+ \pi^+ )}{B(D^0 \rightarrow K^- \pi^+ )} =
2.35 \pm 0.23 \;.
\end{equation}
We can thus express $Y_{D^+}$ in terms of $B(D^0 \rightarrow K^- \pi^+)$ as 

\begin{eqnarray}
Y_{D^+} & = & (0.235 \pm 0.017) \;\frac{9.3\%}{B(D^+ \rightarrow K^-
\pi^+\pi^+)} = \;
\nonumber \\
& = & (0.235 \pm 0.017) \;\frac{9.3\%}{r_+\cdot B(D^0 \rightarrow K^-
\pi^+)} = \;
\nonumber \\
& = & (0.238 \pm 0.029) \;\left[\frac{3.91\%}{B(D^0 \rightarrow K^-  
\pi^+)}\right] \;.
\end{eqnarray}
The inclusive $D$ yield in $\overline B$ decays,
\begin{equation}
Y_D \equiv Y_{D^0} + Y_{D^+} \;,
\end{equation}
can then be expressed in terms of $B(D^0 \rightarrow K^- \pi^+)$ as
shown in Table I. 

The central values for $\Xi_c$ and $\Lambda_c$ yields which are 
typically used in the determination of $n_c$ are both at the 5\% level 
\cite{ncbrowder}. The inclusive $\Lambda_c$ production in 
$B$ decays is measured rather well, whereas the $\Xi_c$ yield has large 
uncertainty. The CLEO experiment has demonstrated that right-sign 
$\ell^+\Lambda_c$ correlations dominate over the wrong-sign 
$\ell^-\Lambda_c$ case \cite{glasgowbaryon}, (where the lepton comes from the
semileptonic decay of one $B$ and the $\Lambda_c$ originates from the other 
$B$ in an $\Upsilon (4S)$ event). As a consequence, inclusive $\Xi_c$ 
production in $B$ decays cannot be as large as that of the $\Lambda_c$.
This is shown in the Appendix where we
relate both $\Xi_c$ and $\Omega_c$ production in
$B$ decays to $Y_{\Lambda_c}$ and the ratio
\begin{equation}
r_{\Lambda_c} \equiv \frac{B(\overline B\rightarrow  
\overline\Lambda_c X)}{B(\overline B\rightarrow \Lambda_c X)}\;.
\end{equation}
We neglect $b\rightarrow u$ transitions and use the Cabibbo suppression
factor, $\theta^2 =(0.22)^2$, for charmed baryon production in
$b\rightarrow c\bar us (b\rightarrow c\bar cd)$ versus $b\rightarrow  
c\bar ud^\prime (b\rightarrow c\bar cs^\prime)$. [The prime indicates that the  
corresponding Cabibbo-suppressed mode is included.]  The Appendix  
also parametrizes $s\bar s$ fragmentation from the vacuum, and predicts
\begin{equation}
\frac{Y_{\Xi_c}}{Y_{\Lambda_c}} = 0.38 \pm 0.10 \;,
\end{equation}

\begin{equation}
\frac{Y_{baryon_c}}{Y_{\Lambda_c}}= 1.41 \pm 0.12 \;,
\end{equation}

\begin{equation}
\frac{B(\overline B\rightarrow baryon_c X)}{Y_{\Lambda_c}} = 1.22 \pm  
0.07 \;,
\end{equation}

\begin{equation}
\frac{B(\overline B \rightarrow \overline{baryon_c} \;X)}{Y_{\Lambda_c}} = 0.20
\pm 0.10 \;.
\end{equation}
As discussed in the Appendix,
$\Xi_c$ 
production in $\overline B$  
decay is probably overestimated.
Inclusive $baryon_c$ production thus lies somewhere in the range
\begin{equation}
1 < \frac{Y_{baryon_c}}{Y_{\Lambda_c}} < 1.41 \pm 0.12 \;.
\end{equation}
Variation over this range has negligible effect on the 
value of $n_c$. We therefore use the values given in 
Eqs. (2.7) - (2.10). 
We also prefer not to use the 1994 PDG 
value~\cite{pdg} of $B(\Lambda_c\rightarrow pK^- \pi^+ )
= (4.4 \pm 0.6)$\%, because it relies upon a flawed model of  
baryon production in $B$ decays. We instead follow the 
approach outlined in Ref.~\cite{shipsey} and use 
$B(\Lambda_c\rightarrow pK^- \pi^+ )=(6.0 \pm 1.5)$\%. 
Thus $n_c$ in Eq.~(2.1) can be written :
\begin{eqnarray}
n_c & = & (0.883 \pm 0.038) \;\left[\frac{3.91\%}{B(D^0 \rightarrow  
K^- \pi^+
)}\right] + (0.1211 \pm 0.0096) \;\left[\frac{3.5\%}{B(D_s  
\rightarrow \phi\pi
)}\right] + \nonumber \\
& + & (0.042 \pm 0.008) \;\left[\frac{6\%}{B(\Lambda_c \rightarrow  
pK^- \pi^+
)}\right] + 2B(\overline B\rightarrow (c\bar c) X) \;.
\end{eqnarray}
Inserting the branching fractions in Table II and estimating \cite{bdy}, 
\begin{equation}
B(\overline B\rightarrow (c\bar c)X) = 0.026 \pm 0.004 \;,
\end{equation}
one obtains
\begin{equation}
n_c = 1.10 \pm 0.06\;\;\;\;
\end{equation}
which is below the currently accepted value of $1.18 \pm 0.06$
~\cite{ncbrowder}.

Very recently, the CLEO experiment has completed the direct measurement of  
$B(b\rightarrow c\bar cs^\prime )$ which allows one to use 
the following, alternative expression for the number of charm quarks
per $B$ decay \cite{bdy},

\begin{equation}
\tilde{n}_c = 1- B(b\rightarrow \;{\rm no}\;{\rm charm}) +  
B(b\rightarrow c\bar
cs^\prime ) \;.
\end{equation}
This expression is much less sensitive to
either $B(\stackrel{(-)}{B}\rightarrow baryon_c X)$ or
$B(D^0 \rightarrow K^- \pi^+ )$.
We take $B(b\rightarrow$no charm) to be \cite{bdy},
\begin{eqnarray}
B(b\rightarrow \;{\rm no}\;{\rm charm})
& = & (0.25 \pm 0.10) \;(0.1049 \pm 0.0046) = \nonumber \\ 
& = & 0.026 \pm 0.010.
\end{eqnarray}
The inclusive wrong charm $B$ decay branching fraction is
expressed as \cite{bdy,fwd,bsbsbar}
\begin{eqnarray}
B(b\rightarrow c\bar cs^\prime ) & = & B(\overline B\rightarrow  
\overline D X)
+ B(\overline B\rightarrow D^-_s X) + \nonumber \\
& + & B(\overline B\rightarrow \overline{baryon}_c \;X) + B(\overline
B\rightarrow (c\bar c) X) \;.
\end{eqnarray}
From Tables I and III, Eq.~(2.17) and the charmed baryon correlations 
discussed in the Appendix, we thus obtain
\begin{eqnarray}
B(b\rightarrow c\bar cs^\prime ) & = & (0.085 \pm 0.025)  
\;\frac{3.91\%}{B(D^0
\rightarrow K^- \pi^+ )}  \nonumber \\
& + & (0.100 \pm 0.012) \;\left[\frac{3.5\%}{B(D_s \rightarrow  
\phi\pi
)}\right] +(0.0059 \pm 0.0031) \;\left[\frac{6\%}{B(\Lambda_c  
\rightarrow pK^-
\pi^+ )}\right]  \nonumber \\
& + & B(\overline B\rightarrow (c\bar c) X) \;.
\end{eqnarray}
Using the absolute charm branching fractions of Table II we obtain
\begin{equation}
B(b\rightarrow c\bar cs^\prime )  = 0.22 \pm 0.03 \; ,
\end{equation}
\begin{equation}
\tilde{n}_c = 1.19 \pm 0.03 \;.
\end{equation}
The quantities $n_c$ and $\tilde{n}_c$ must be equal. Their difference 
can be traced to a significant discrepancy in two alternative determinations of
$B(\overline{B}\rightarrow DX)$. On the one hand, one can write
\begin{eqnarray}
B(\overline B\rightarrow DX) & = & 1- B(\overline B\rightarrow \;{\rm  
no}\;{\rm charm}) -
B(\overline B\rightarrow D^+_s X) + \nonumber \\
& - & B(\overline B\rightarrow baryon_c X) - B(\overline B\rightarrow  
(c\bar c ) X) \;.
\end{eqnarray}
Inserting the values from Eqs.~(2.13) and (2.16) and the current CLEO
results,
\begin{equation}
B(\overline B\rightarrow D^+_s X) = (0.021 \pm 0.010)  
\left[\frac{3.5\%}{B(D_s \rightarrow
\phi\pi )}\right] \;,
\end{equation}

\begin{equation}
B(\overline B\rightarrow baryon_c X) = (0.0365 \pm 0.0065  
)\left[\frac{6\%}{B(\Lambda_c
\rightarrow pK^- \pi^+ )}\right] \;,
\end{equation}
into the right hand side of Eq.~(2.21) we obtain 
\begin{equation}
B(\overline B\rightarrow DX) = (0.89\pm0.02)~~.
\end{equation}
On the other hand, current CLEO measurements of $Y_D$ and $r_D$
(see Tables I and III) yield,
\begin{equation}
B(\overline{B} \rightarrow DX)=
(0.798 \pm 0.042) \left[\frac{3.91\%}{B(D^0 \rightarrow
K^-\pi^+ )}\right] \;.
\end{equation}
Equating the two determinations of $B(\overline{B}\rightarrow DX)$,
\begin{equation}
(0.89 \pm 0.02) = (0.798 \pm 0.042) \left[\frac{3.91\%}{B(D^0 \rightarrow
K^-\pi^+ )}\right] \;,
\end{equation}
it follows that 
either the coefficient $(0.798 \pm 0.042)$, or $B(D^0 \rightarrow K^-\pi^+ )$
$=~(3.91\pm 0.19)\%$, or both, are incorrect.  Let us assume for the moment
that only $B(D^0 \rightarrow K^-\pi^+ )$  is incorrect. We can then
solve Eq.~(2.26) for $B(D^0 \rightarrow K^-\pi^+ )$ to obtain,
\begin{equation}
B(D^0 \rightarrow K^- \pi^+) = (3.50 \pm 0.21)\% \; .
\end{equation}
This is
considerably smaller than currently accepted values but compatible with the 
most recent measurement from ARGUS\cite{argusd0},
$$B(D^0 \rightarrow K^- \pi^+ )~=~(3.41\pm0.12\pm0.28)\%.$$
Eq.~(2.27), in turn, yields
\begin{equation}
B(b\rightarrow c\bar cs^\prime )  = (22.7 \pm 3.5)\% \; ,
\end{equation}
\begin{equation}
n_c = \tilde{n}_c = 1.20 \pm 0.04 \;.
\end{equation}

Our result must of course be corroborated by additional precision studies. 
In the meantime we have investigated some consequences of a lower value
for $B(D^0\rightarrow K^-\pi^+)$.

\vskip 0.5 in
\begin{large}
\subsection{The Low $R_c$ Measurement}
\end{large}

Whereas theory predicts
\begin{equation}
R_c \equiv \frac{\Gamma (Z^o \rightarrow c\bar c)}{\Gamma (Z^o
\rightarrow
\;{\rm hadrons})} = 0.172 \;,
\end{equation}
experiments yield a combined result which is $\sim2 \sigma$  
lower \cite{graziani}
\begin{equation}
R_c |_{exp} = 0.1598 \pm 0.0069 \;.
\end{equation}

To analyze this result one must make distinctions
among the various contributing measurements. 
Those which fully reconstruct a primary $D^{*+}$ are
calibrated by $B(D^o \rightarrow K^- \pi^+)$. These are \cite{delphirc,behnke}

\begin{equation}
R_c (DELPHI \;D^*) = 0.148 \pm 0.007 \pm 0.011 \;,
\end{equation}
\begin{equation}
R_c (OPAL\;D^*) = 0.1555 \pm 0.0196 \;,
\end{equation}
with a world-average of
\begin{equation}
R_c (D^*)= 0.150 \pm 0.011 \;.
\end{equation}

Unfortunately, the uncertainty in $R_c$ measurements due to 
$B(D^o\rightarrow K^- \pi^+)$ has not been explicitly reported.
We therefore conservatively retain the full error on $R_c$ to write
\begin{equation}
0.172 = 
(0.150\pm 0.011)\left[\frac{3.84\%}{B(D^o\rightarrow K^- \pi^+)}\right]\;.
\end{equation}
(Note that the calibration factor is different than that used previously
because these measurements have taken the updated PDG value~\cite{pdg95} 
$B(D^o\rightarrow K^- \pi^+)~=~(3.84\pm 0.13)\%$.)
This yields
\begin{equation}
B(D^o \rightarrow K^-\pi^+) =  (3.35 \pm 0.25)\% \; ,
\end{equation}
which is near to the value we extracted in Eq.~(2.27).

Note that DELPHI has also measured $R_c$ via an inclusive double tag method, 
where only the daughter pion of the $D^{*\pm}$ is identified.
This method does not involve $B(D^o \rightarrow K^- \pi^+)$ and
the result \cite{graziani,delphirc}, albeit of limited precision,
\begin{equation}
R_c (\pi^+\pi^-) = 0.171^{\textstyle{+0.014}}_{\textstyle{-0.012}}
\pm 0.015
\end{equation}
agrees well with the Standard Model.

Other measurements of $R_c$ include a lepton method which has very large 
systematic uncertainties, and measurements from both OPAL and DELPHI 
that involve direct charm counting \cite{graziani,opalrc}. The extraction of 
$B(D^o \rightarrow K^- \pi^+)$ from the latter is less straightforward 
since a variety of charmed hadrons are involved. Consider, for instance,
the recent OPAL result \cite{opalrc},

\begin{equation}
R_c ({\rm charm}\;{\rm counting}) =  0.167 \pm 0.011 (stat) \pm
0.011\;(sys) \pm 0.005\;(br)\;.
\end{equation}
OPAL measures
\begin{eqnarray}
R_c \;f(c\rightarrow D^o)\;B(D^o \rightarrow K^- \pi^+) & = & (0.389
\pm 0.037) \%
\;,\nonumber \\
R_c \;f(c\rightarrow D^+) \;B(D^+\rightarrow K^- \pi^+\pi^+) & = &
(0.358 \pm 0.055) \%
\;, \nonumber \\
R_c \;f(c\rightarrow D^+_s) \;B(D^+_s \rightarrow \phi\pi^+) & = &
(0.056 \pm 0.017) \%
\;, \nonumber \\
R_c \;f(c\rightarrow \Lambda^+_c ) \;B(\Lambda^+_c \rightarrow pK^-
\pi^+ ) & = &
(0.041 \pm 0.020) \% \;.
\end{eqnarray}
The fractions are summed by using the updated PDG
branching fractions \cite{pdg95} as reference:
\begin{eqnarray}
B(D^o \rightarrow K^- \pi^+) & = & (3.84 \pm 0.13)\% \;, \nonumber \\
B(D^+\rightarrow K^- \pi^+ \pi^+ ) & = & (9.1 \pm 0.6)\% \;,
\nonumber \\
B(D^+_s \rightarrow \phi\pi^+) & = & (3.5 \pm 0.4)\% \;, \nonumber \\
B(\Lambda_c \rightarrow pK^- \pi^+) & = & (4.4 \pm 0.6)\% \;.
\end{eqnarray}
They assume that the undetected primary $\Xi_c$ and $\Omega_c$
production is $(15 \pm 5) \%$ of the primary $\Lambda_c$ production,
and thus obtain Eq.~(2.38).

We assume the standard model value $R_c = 0.172$ and again use 
the more accurate estimate for
$B(\Lambda_c\rightarrow pK^- \pi^+)$ of $(6.0 \pm 1.5)\%$, rather
than $(4.4 \pm 0.6)\%$, in order to solve for $B(D^o \rightarrow K^- \pi^+)$.
We correlate the inclusive primary production fraction of
$baryon_c$ to that of $\Lambda_c$ via
\begin{equation}
f(c\rightarrow baryon_c ) =  \;f(c\rightarrow \Lambda_c )\;/ \;
(1-p)^2\;,
\end{equation}
where $p$ models the production fraction of $s\bar s$ fragmentation
relative to $f\bar f$ fragmentation from the vacuum, ($f=u, d$ or $s$)
\cite{pvalue}.
The solution for $B(D^o \rightarrow K^- \pi^+)$ is 
\begin{equation}
B(D^o \rightarrow K^- \pi^+) = \frac{(0.00389 \pm 0.00037 ) + \frac{0.00358
\pm 0.00055}{r_+}}{R_c - \frac{(0.00056 \pm 0.00017)}{B(D_s
\rightarrow \phi\pi^+)} - \frac{(0.00041 \pm 0.00020)}{(1-p)^2
B(\Lambda_c\rightarrow
pK^- \pi^+ )}}\; .
\end{equation}
Inserting,
\begin{equation}
r_+ = 2.35 \pm 0.23, ~~R_c = .172, ~~B(D_s \rightarrow \phi\pi^+) =
(3.5 \pm 0.4)\% \;,
\end{equation}

\begin{equation}
{\rm and}\;B(\Lambda_c \rightarrow pK^- \pi^+ ) = (6.0 \pm 1.5 )\%,
\end{equation}
we obtain
\begin{equation}
B(D^o \rightarrow K^- \pi^+) = (3.67 \pm 0.36)\% \;.
\end{equation}

\vskip 0.5 in
\begin{large}
\subsection{Semileptonic $B$ Decays}
\end{large}

Semileptonic $B$ transitions are among the most intensively studied $B$
decays. They consist almost entirely of $b\rightarrow c\ell^- \bar\nu$
transitions, since $|V_{ub}/V_{cb}| \approx 0.1\;.$
Thus a primary lepton in $B$ decay is typically accompanied by a charmed hadron.
Inclusive semileptonic $B$ decay measurements detect the
lepton without reconstructing the accompanying charmed
hadron. As a result, uncertainties from charm are minimal. 
These measurements also usually involve very high statistics and so they 
are generally very precise~\cite{lepew96,burchatr,cleoslincl,lepsl}.

A variety of semileptonic $B$ decay measurements, where the accompanying charm 
was also seen, have been reported~\cite{burchatr}. These include the
dominant exclusive $\overline B\rightarrow D^{(*)} \ell^-\bar\nu$
processes,
$\overline B\rightarrow D^{**} (X) \ell^- \bar\nu$ transitions, and
non-resonant
$\overline B\rightarrow D^{(*)} \pi X\ell^- \bar\nu$ processes.
Combining all information about semileptonic $B$ decay
measurements where the associated charm is also seen, one finds  
a significant shortfall relative to the
inclusive measurements \cite{drellp}. Decreasing $B(D^0 \rightarrow
K^- \pi^+)$ would alleviate this shortfall, because the semileptonic
branching fractions with reconstructed charm are
inversely proportional to $B(D^0\rightarrow K^-\pi^+)$ and 
would therefore increase. With some theoretical input we estimate 
that the value $B(D^0\rightarrow K^-\pi^+)$ $=~(2.9\pm 0.4)\%$ 
eliminates the discrepancy \cite{recal}.

\vskip 0.5 in
\begin{large}
\subsection{Summary and Implications}
\end{large}

We have demonstrated that 
currently accepted values for $B(D^0 \rightarrow K^- \pi^+)$, namely

\begin{eqnarray}
B(D^0 \rightarrow K^- \pi^+) = \left\{\begin{array}{cc}
(4.01 \pm 0.14)\% & (1994\; PDG) \\
(3.91 \pm 0.19)\% & (CLEO\; II) \\ 
(3.84 \pm 0.13)\% & (1995\; PDG \; update ) \; 
\end{array}\right.
\end{eqnarray}
could be too high.
The recent wrong charm $B(\overline B\rightarrow \overline DX)$
measurement of CLEO opened up a second way to determine 
right charm $B(\overline{B} \rightarrow  DX).$ 
By equating the two determinations, we solved for $B(D^0
\rightarrow K^- \pi^+)$ to obtain the smaller value,
\begin{equation}
B(D^0 \rightarrow K^- \pi^+) = (3.50 \pm 0.21)\% \;.
\end{equation}
We then demonstrated that reducing the value of $B(D^0\rightarrow K^- \pi^+)$
enables experimental results for $R_c$ 
to agree with theory and diminishes the excess of inclusive
semileptonic $B$ decays relative to the combined exclusive yields.
Table \ref{table:dkpi} lists the values of $B(D^0\rightarrow~K^-\pi^+)$
required to eliminate the discrepancy in each of these cases.
Combining these values one obtains the weighted mean value: 
$$\langle B(D^0 \rightarrow K^- \pi^+)\rangle = (3.40 \pm 0.14)\% \;.$$

Additional consequences of a lower value for $B(D^0 \rightarrow K^- \pi^+)$
are discussed in Ref. \cite{recal}.  We note here that it is possible for
a reduction in $B(D^0 \rightarrow K^- \pi^+)$ to affect 
the discrepancy between theory and measurement for $R_b$, the
fraction of hadronic Z decays to bottom quarks. This connection 
follows by noting that since $B(D^0 \rightarrow K^- \pi^+)$ calibrates 
almost all charmed meson branching fractions, a lower value for
$B(D^0 \rightarrow K^- \pi^+)$ means 
that a significant fraction of $D^0$ and $D^+$ decays have not been 
observed or properly counted. One hypothesis is that these missed decays
involve high track multiplicities\cite{appel} since such decays are more 
difficult to fully reconstruct due to tracking inefficiencies, particle 
identification errors, combinatoric backgrounds and 
the presence of undetected neutrals. On 
the other hand, high charged multiplicity decays are 
more likely to generate a lifetime $B$ tag at LEP and SLC since they will
more likely yield the high number of significantly displaced tracks 
expected for $B$ decays. We explore this possibility further in the
next section.

\vskip 0.5 in
\begin{large}
\section{\Rb Measurements}
\end{large}

Recently the fraction $R_b$ of Z hadronic decays to $\bbbar$ has been
measured at LEP \cite{ashape,lshape,alife,dmultiple,omultiple} and
SLC \cite{sld} using a variety of methods including shape
variables, multivariate techniques, high $p_T$ leptons, and lifetime tags
to distinguish the decays of b quarks from those of lighter quarks.
While each measurement is consistent, within uncertainties, with the
Standard Model expectation of $R_b~=~0.2155$, they combine
to yield $R_b~=~0.2205\pm0.0016$\cite{cernppe95172}
which represents a three standard deviation discrepancy.
As seen in Table \ref{table:Rb}, the highest precision
contributions to this average are those which use lifetime $B$ tagging.
Indeed, the lifetime measurements (including the lepton $+$ lifetime
result from OPAL) yield a simple
weighted mean value of $R_b~=~0.2200\pm0.0017$
which dominates the overall result.

The procedure used for measuring $R_b$ is to tag
$Z \rightarrow \bbbar$ events using any of the above-mentioned methods, then
subtract backgrounds as estimated
from Monte Carlo (MC), and estimate the $B$ tag efficiency either by MC
or directly from data. Obtaining the $B$ tag efficiency from data is more 
reliable and is possible in all cases where
double tagging (tagging two $B$ hadron decays in one event) is used.
As an illustration of the procedure, if we were to ignore backgrounds,
the number of tagged hemispheres $N_t$, (where the sphere axis is defined by
the direction of the highest energy jet)
and the number of double tagged events $N_{tt}$ would be expressed as
\begin{equation}
N_t = 2 \ebi R_b N_Z
\end{equation}
\begin{equation}
N_{tt} = C_b \ebi^2 R_b N_Z  
\end{equation}
For a given $B$ tagging algorithm one counts $N_t$ and $N_{tt}$. $N_Z$ is the
total number of hadronic Z decays. $C_b$ is a correlation factor which takes 
into account the fact that the probability of tagging a hemisphere may be
correlated with whether or not the other hemisphere is
tagged. Eqs. (3.1) and (3.2) can be solved for \Rb and \eb :
\begin{equation}
\ebi = {2N_{tt}\over C_b N_Z} 
\end{equation}
\begin{equation}
R_b = {C_b N_t^2\over 4N_{tt}N_Z} 
\end{equation}
When the $B$ purity of the tagging algorithm is high, so that backgrounds are
small, Eqs. (3.3) and (3.4) are relatively good approximations.
Including backgrounds, one arrives at the following generalizations of
Eqs. (3.1) and (3.2):
\begin{equation}
{N_t \over 2N_Z} = \ebi R_b +\eci R_c +\eudsi (1- R_b - R_c) 
\end{equation}
\begin{equation}
{N_{tt}\over N_Z} =  C_b\ebi^2 R_b  +C_c\eci^2 R_c +
C_{uds}\eudsi^2 (1-R_b-R_c) 
\end{equation}
As input to these equations one can use either the SM value $R_c~=~0.172$
or a value measured in a separate analysis, and MC
predictions for \ec and \euds. A variety of data
and MC-based studies are performed to estimate the correlation factor
$C_b$. $C_c$ and $C_{uds}$ differ negligibly from unity.

We have asked whether there could be an explanation for the experimental
result which does not contradict the Standard Model.
Under the $assumption$ that there could be an error in the result obtained
for the dominant lifetime measurement technique, it follows that 
this error would have to be common to a
number of different measurements and, as indicated by
equation (3.4), would likely result from an excess of single 
hemisphere tags relative to double-tagged events, or from an overestimate
of $C_b$. Due to the many subtle differences in the experiments and analyses,
it is unlikely that a common error could have occurred in the
estimates of $C_b$. There could however be a missed or
incorrectly estimated background that enhances
$N_t$ and/or suppresses $N_{tt}$. Since the $b$ tagging algorithms
used by the experiments are of high purity, as shown in Table \ref{table:btag},
we conclude that events with two tagged hemispheres should have
extremely small non-b backgrounds. By process of elimination, one is
therefore led to the possibility that
some non-b source of single hemisphere tags could have been overlooked.
From Eqs.~(3.4) and (3.5) this would imply that one or 
both of \ec and \euds are wrong.

The light quark tagging efficiency \euds 
can be studied directly with data even for
the added complication of gluon radiation with splitting
to $\bbbar$ or $\ccbar$\cite{lepglue}. The charmed tagging efficiency \ec is
more difficult to determine since it depends upon
the relative populations of $D^o$, $D^+$,
$D_s$, $\Lambda_c$ ...,  and their decays.
We can investigate the sensitivity of $R_b$ to \ec and \euds by plotting
the variation in $R_b$ as a function of 
each efficiency with all other quantities fixed
at their nominal values. For simplicity we set $C_b=1$ which then results in
the following simple expression for $R_b$:

\begin{equation}
R_b = { ({N_t \over 2 N_Z} - R_c(\eci-\eudsi)-\eudsi)^2 \over
{N_{tt}\over N_Z} - {N_t \over N_Z}\eudsi - R_c(\eci-\eudsi)^2+\eudsi^2}
\end{equation}

\begin {figure} [tbhp]
   \epsfysize=6.0 in
\center{\leavevmode
\epsffile[54 144 522 648]{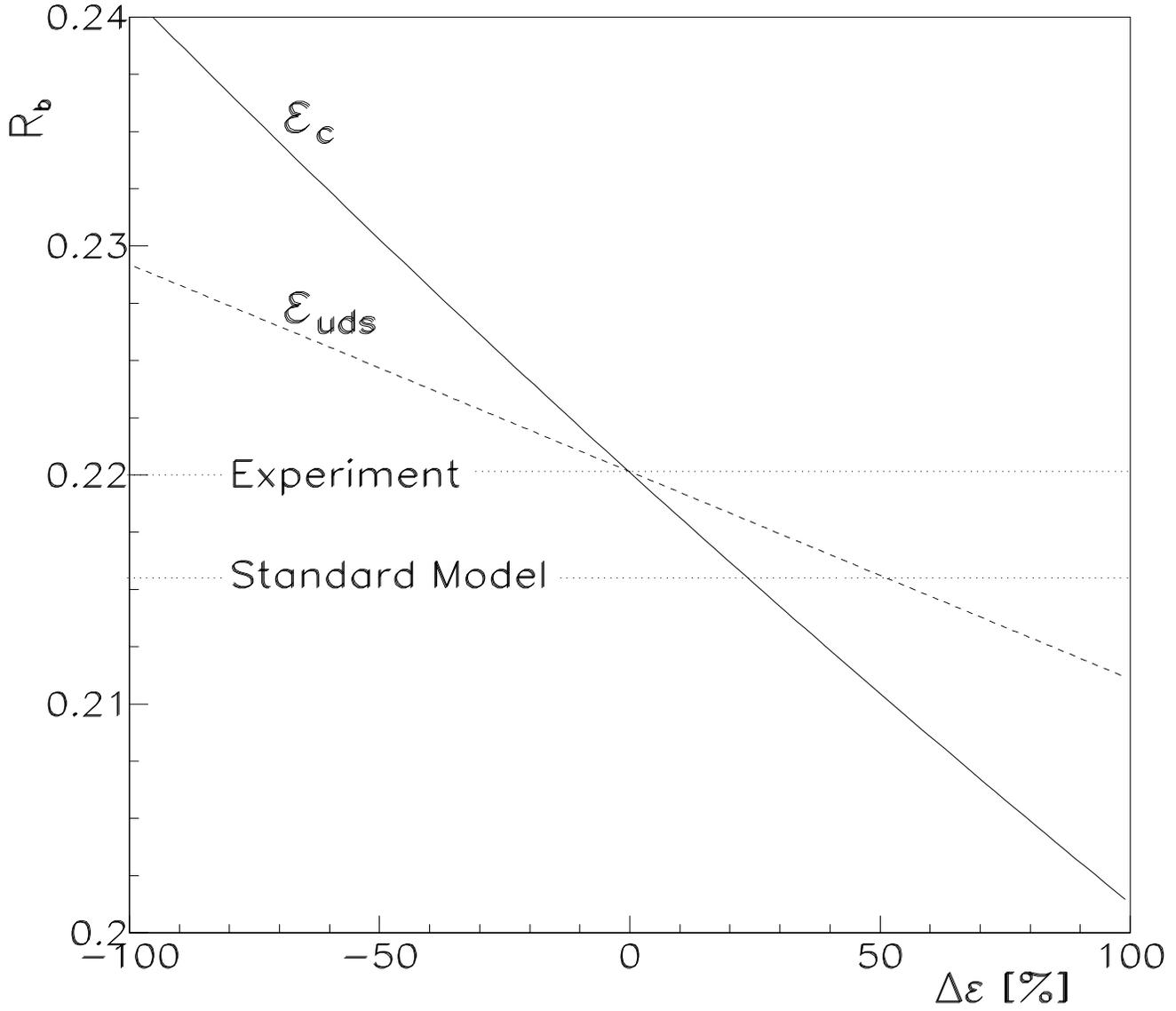}}
\vspace{4.0cm}
\caption{ \Rb as a function of the change in the efficiencies
\euds and \ec varied separately. }
\label{fig:rb_eps}
\end {figure}

For nominal values we choose $\eudsi = 1.7\times10^{-3}$ and
$\eci = 1.4\times10^{-2}$ which are in the mid-range of the values
used by the LEP experiments (see Table \ref{table:btag}). We assume the value
$R_c=0.172$. Finally
we use ${N_t \over 2 N_Z}~=~0.055$ and ${N_{tt}\over N_Z}~=~0.012$ 
which are comparable to the values in Ref. \cite{omultiple} and result in
$R_b~=~0.22$ as quoted above for the combined lifetime
measurements. Figure \ref{fig:rb_eps} plots $R_b$ as a function of the 
change in \ec and \euds. 
The discrepancy between measured and
predicted values for $R_b$ is seen to correspond to roughly a
$50 \%$ increase in \euds from its nominal value or
a $20\%$ increase in \ec. (The discrepancy is of course also
removed by smaller but simultaneous upward shifts in both efficiencies.)

A review of the various measurements of \euds yields
no obvious oversights. Our initial concern was that
there could be some contribution from gluon radiation followed by splitting
to $\bbbar$ or $\ccbar$. However, $g\rightarrow\bbbar$ occurs
in only about (0.2~-~0.3)$\%$ of Z hadronic decays  while
$g\rightarrow\ccbar$ occurs in roughly 2.5$\%$ of Z hadronic
decays \cite{lepglue,gtheory1,gtheory2,gtheory3}.
Furthermore, the experiments have explicitly studied the
effect of $g\rightarrow\bbbar,\ccbar$ in Z decays to light quarks and find
no significant enhancement in \euds. Even under
the assumption of a large uncertainty in the probability for this phenomenon
to occur, it is not possible to achieve anywhere near the 50$\%$
shift required in \euds to explain the $R_b$ discrepancy.

We next consider \ec. We were unable to find explicit
consideration of the possible effect of gluon radiation and splitting to
$\bbbar$ and $\ccbar$ in $Z\rightarrow \ccbar$ events. The effect of gluon
splitting in these events could be more important than
in the light quark Z decays since: $(i)$ the $R_b$ discrepancy is removed by a 
smaller change in \ec than would be required
for \euds and $(ii)$ $Z\rightarrow\ccbar g$ with $g\rightarrow\ccbar$
results in events with typically three heavy flavors in one hemisphere. Such
hemispheres could be expected to have a higher than average multiplicity of
significantly displaced charged tracks.  The primary
charm quark from the Z decay will typically remain fairly energetic
in spite of having radiated a gluon, and it is then
only necessary for one of the charm quarks from $g\rightarrow\ccbar$
to be energetic in order to have two significantly displaced heavy flavor
decays similar to that expected for $B$ decay.

To study this phenomenon we have used the
PYTHIA Monte Carlo (version 5.7) with JETSET (version 7.4) \cite{pythia}.
To cross-check these results we have written a toy MC based upon the
differential relative rate for the process $Z^0 \to c \overline{c}
g,\; g \to c \overline{c}$ calculated using an
${\cal O}(\alpha_s^2)$
cross section.  Our calculation is more
precise than the leading-log approximation at ${\cal O}(\alpha_s^2)$
valid only at small angles, because we wanted to also consider
large-angle gluon emission.  We neglected terms
proportional to $k^2/M_Z^2$, with $k^2$ denoting the virtuality of
the gluon.
As anticipated, we find that the fraction of those hemispheres with
$g\rightarrow\ccbar$ that contain a large number of displaced charged tracks
(e.g. $N_{displaced} \ge 4$) increases dramatically. These however represent
only a small fraction of all hemispheres with $g\rightarrow\ccbar$ and
upon applying the jet probability $B$ tag algorithm 
\cite{alife} we found that, on average, the tagging efficiency was not 
significantly higher than for hemispheres which do not contain 
$g\rightarrow\ccbar$. 

Finally, as mentioned in the previous section, 
the accepted value for the branching ratio $B(D^0 \rightarrow K^-\pi^+)$ 
could be too high. One consequence would be that roughly $15 \%$ of 
$D^0$ decays would be missing. A similar argument applies to $D^+$ decays 
since the most accurate measurement of $B(D^+\rightarrow K^-\pi^+\pi^+)$
is tied to $B(D^0\rightarrow K^-\pi^+)$ \cite{cleorp}.
A number of different scenarios could be proposed for the 
missing decays. We consider here one which
could link the reduced value for $B(D^0 \rightarrow K^-\pi^+)$ to
the discrepancy between the measured value of $R_b$ and the value
predicted by the Standard Model. In particular, if the missing
decays were predominantly multiple charged particle modes, then
it can be demonstrated that this would increase the value of \ec relative
to that currently assumed in recent measurements of $R_b$.

To demonstrate this we have performed a simple Monte
Carlo study of the process $e^+e^-\rightarrow Z\rightarrow\ccbar$, again 
using the PYTHIA MC with JETSET.
We have chosen the DELPHI detector for the purpose of 
creating a simple model of detector effects such as silicon detector 
acceptance and impact parameter resolution. 
The results would however be qualitatively the same if we were
to use OPAL or ALEPH detector acceptances and resolutions.
For each stable charged particle we have calculated the impact
parameter $(d)$  relative to the $e^+e^-$ interaction vertex. The 
calculated impact parameter is smeared according to the measured 
DELPHI resolution function \cite{graziani}. The impact parameter significance
is then defined as $S\equiv d/\sigma_d$ where $\sigma_d$ 
depends on transverse momentum \cite{physrep}.

\begin {figure} [tbhp]
   \epsfysize=6.0 in
\center{\leavevmode
\epsffile[54 144 522 648]{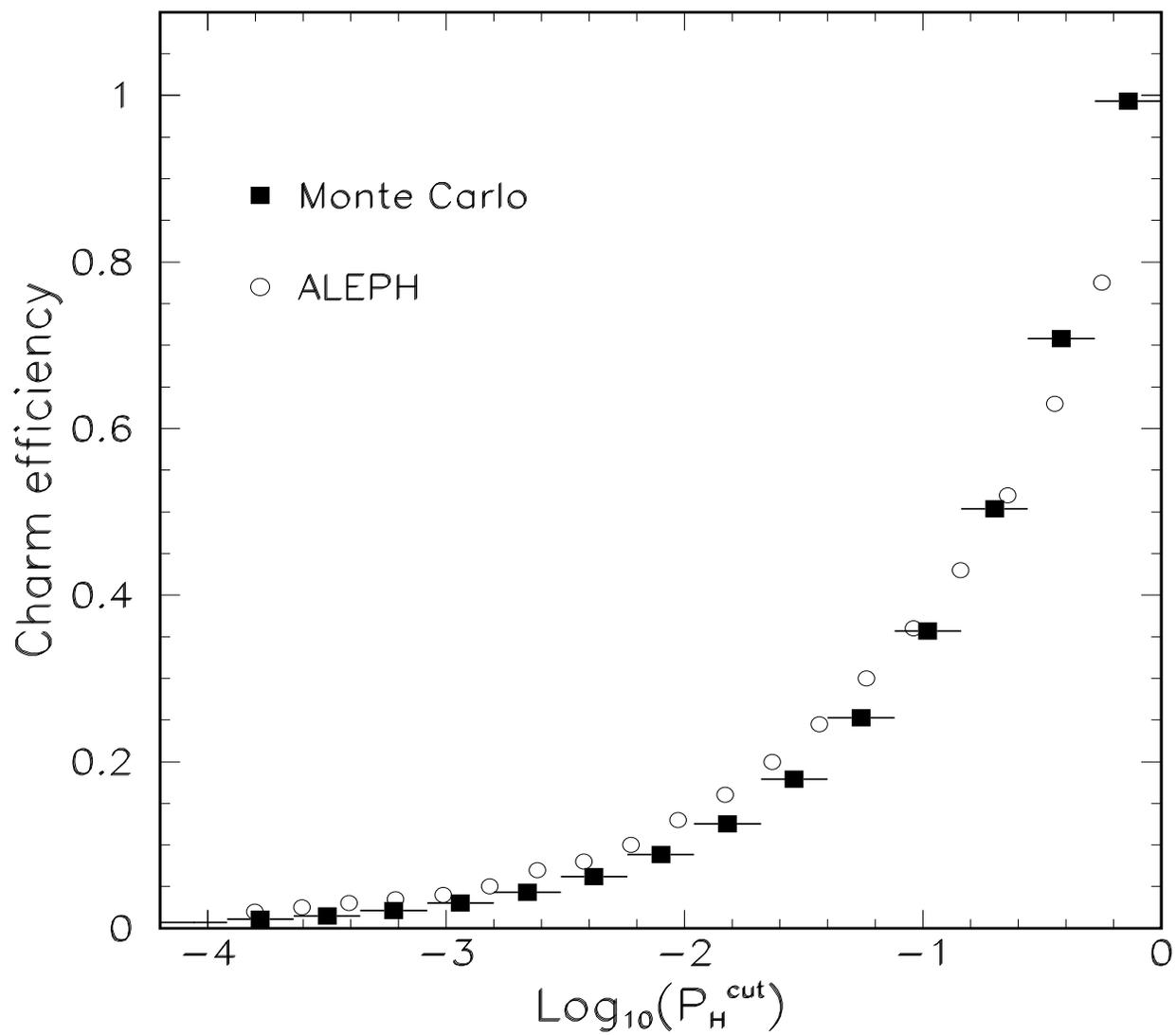}}
\vspace{4.0cm}
\caption{ The tag efficiency versus the $\rm{Log}$ of the probability cut
($\rm{Log_{10}(P_H)}$) for Monte Carlo
$e^+e^-\rightarrow Z\rightarrow\ccbar$ events used in this paper
(open circles) as compared with that obtained by ALEPH (solid squares)}.
\label{fig:jpb}
\end {figure}

\begin {figure} [tbhp]
   \epsfysize=6.0 in
\center{\leavevmode
\epsffile[54 144 522 648]{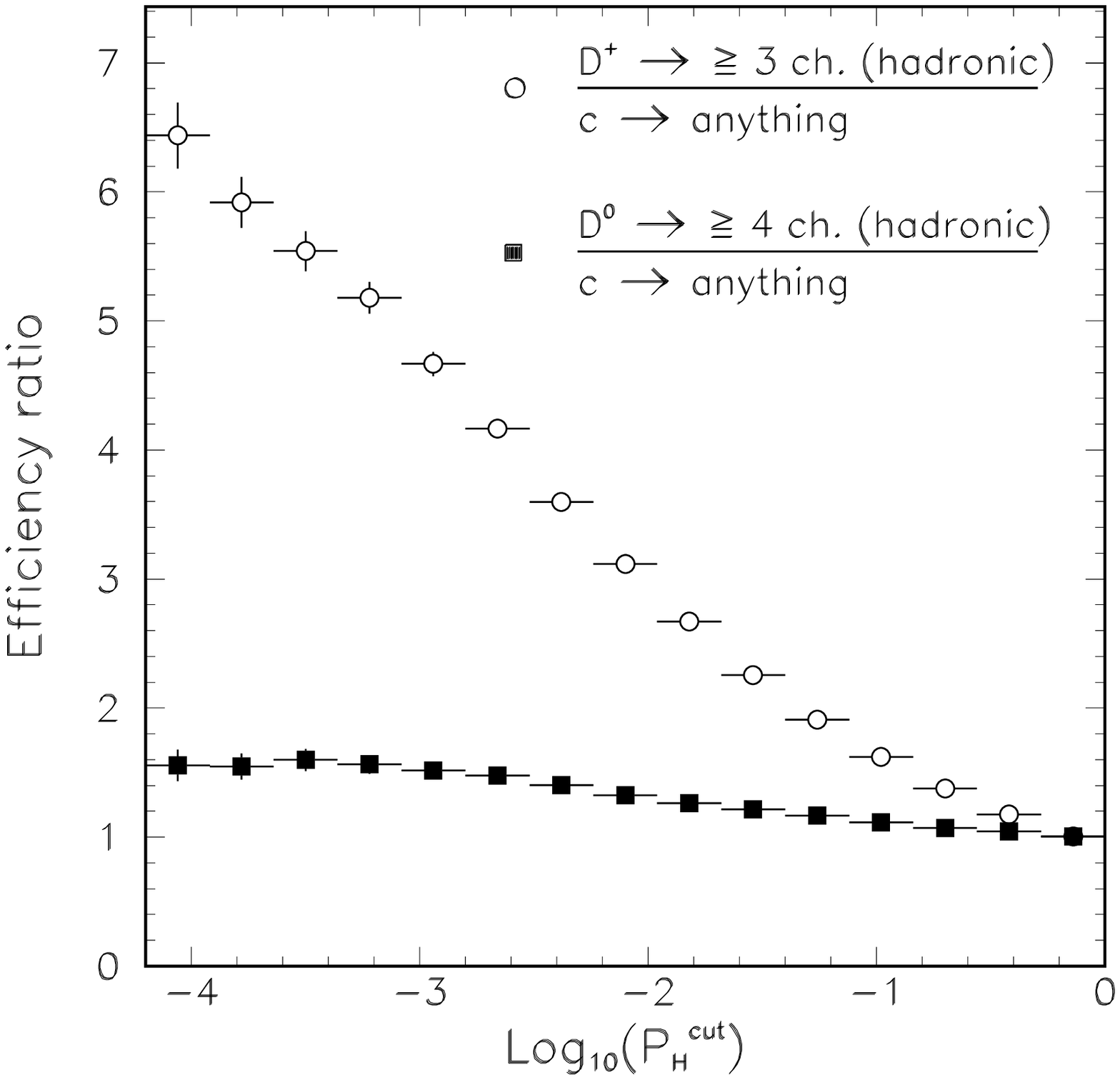}}
\vspace{4.0cm}
\caption{ The ratio of the tagging efficiency for $D^0$ hadronic
decays containing at least 4 charged particles to the decays
of all charmed hadrons (black squares) and the similar ratio 
for $D^+$ hadronic decay modes containing at least 3 charged particles
(open circles) as a function of $\rm{Log_{10}}$ of the
cut on hemisphere jet probability $\rm{P_H^{cut}}$.  }
\label{fig:highmult}
\end {figure}

For lifetime $B$ tagging we again apply the jet probability algorithm 
which uses impact parameter significance values of charged particle 
tracks to detect the presence of long-lived 
particles. The resulting hemisphere probability distribution we obtain for 
$e^+e^-\rightarrow Z\rightarrow\ccbar$ is shown in Figure \ref{fig:jpb}
where it is compared to that obtained by ALEPH \cite{alife}. The distributions
are not expected to be in strict agreement since we have not 
modelled detector efficiencies and acceptances in detail. Nevertheless, the
agreement is good and should be adequate for this discussion.
Figure \ref{fig:highmult} plots the hemisphere tagging
efficiencies for all $D^0$ decays containing 4 or more charged 
particles, and for all $D^+$ decays containing 3 or more charged
particles (normalized to the tagging efficiency for all charm
hemispheres), as a function of
$\rm{log_{10}(P_H^{cut})}$, where $\rm{P_H}$ is the jet probability 
obtained for all charged tracks in the hemisphere. In measurements of
$R_b$ using this algorithm, a $B$ tag is defined as a hemisphere 
satisfying $\rm{P_H~\le~P_H^{cut}}$ where 
$10^{-4} \le \rm{P_H^{cut}} \le 10^{-2.9}$ 
for the various experiments. It is clear from Figure \ref{fig:highmult} that
the multiple charged track modes have a significantly higher
value of \ec in this range of $\rm{P_H^{cut}}$ values. 
Despite the presence of fewer charged tracks,  the $D^+$ efficiency
is much higher than that of the $D^0$ in Figure \ref{fig:highmult}, 
becaue of its significantly longer lifetime.

From our previous discussion, we found that the $R_b$ discrepancy is removed
by a $20\%$ increase in \ec. Relative to a nominal value of 
$\eci=0.014$, this corresponds to $\eci=0.017$. Since the charm quark
hadronizes as a $D^0$ or $D^+$ roughly 60\% and 25\% of the time, respectively,
the $\sim15\%$ of missing decays corresponds to $\sim13\%$ of all charm 
hemispheres. Let us assume that the remaining $\sim87\%$ of charm
hemispheres have efficiency $\eci=0.014$. We will take
$\rm{P_H^{cut}~=~10^{-3.5}}$ and assume that all
missing decays are high multiplicity decays for which the
efficiencies are given by Figure \ref{fig:highmult}. This 
allows us to estimate the maximum impact on $R_b$. 
From Figure \ref{fig:highmult} at $\rm{P_H^{cut}~=~10^{-3.5}}$, 
$D^+$ decays to three or more charged particles have a tagging
efficiency $\eci^+~=~5\cdot\eci$ while $D^0$ decays to four or more 
charged particles have efficiency $\eci^0~=~1.5\cdot\eci$.
The adjusted charm efficiency is therefore given by:
$$\eci'=0.87\cdot\eci + 0.15\cdot(0.6\cdot\eci^0 + 0.25\cdot\eci^+)= \\
1.20\cdot\eci= 0.017~$$
Thus, in the extreme where all missing decays are multiple charged particle
modes, the $R_b$ discrepancy is completely eliminated.
We hasten to add that, in reality, charged modes may not represent all of the 
missing decays so that the effect on $R_b$ could be smaller. In fact, 
at the opposite extreme, if the missing decay modes involve few
or no charged particles this would lead to an increase in the
final value of $R_b$. 

\vskip 0.5 in
\begin{large}
\section{Conclusions}
\end{large}

We have demonstrated 
that complementary determinations of $B(\overline{B}\rightarrow DX)$
can be reconciled by a downward revision in the 
branching ratio $B(D^0\rightarrow K^-\pi^+)$. This revision, 
together with the central
role played by $B(D^0\rightarrow K^-\pi^+)$ in calibrating almost all
charmed hadron yields, explains 
the $R_c$ discrepancy and diminishes a problem in semileptonic $B$ decays.
We have speculated that it could also be linked to the $R_b$ puzzle.
In particular, a reduction in $B(D^0\rightarrow K^-\pi^+)$ would mean that
roughly 15\% of all $D^0$ and $D^+$  decays have not been properly seen or
counted. In the case where all of the missing decays involve multiple charged 
particle final states, we demonstrated that this leads to a higher than
anticipated lifetime tag efficiency in $Z\rightarrow \ccbar$ events which
would be adequate to bring the measured value of $R_b$ into line with
the Standard Model value of 0.2155~. We have also explicitly considered
$g\rightarrow \ccbar$ in $Z\rightarrow \ccbar$ events but find that it does
not enhance the lifetime tagging efficiency.

We would like to thank M.S.~Alam, J.~Appel, T.~Behnke, M.~Beneke, K.~Berkelman,
D.~Bloch, T.E.~Browder, G.~Buchalla, D.~Charlton, 
P.S.~Cooper, J.~Cumalat, A.S.~Dighe, L.~Dixon, P.~Drell,
E.~Eichten, R.K.~Ellis, R.~Forty, A.~Freyberger, D.~Fujino, W.~Giele,
E.~Graziani, N.~Isgur, V.~Jain, R.~Kowalewski, J.~K\"uhn, R.~Kutschke, Y.~Kwon,
A.J.~Martin, M.~Morii, R.~Poling, J.~Richman, T.~Riehle, P.~Roudeau,
A.~Ryd, D.~Scora, F.~Simonetto, R.~Tenchini, E.~Thorndike,
S.~Veseli, R.~Wang, B.~Winstein, and H.~Yamamoto for discussions.
We thank R. Forty for providing us with relevant Ph.D. theses of the ALEPH
collaboration.  This work was supported in part by the Department
of Energy, contract number DE-AC02-76CH03000.

\vskip 0.5 in
\begin{large}
\section{Appendix: Charmed Baryon Production in $B$ Meson Decays}
\end{large}

Accurate accounting of inclusive charm yields in $B$ decays requires a
consistent description of charmed baryon production, which is lacking in the 
existing literature. Several years ago it was hypothesized that the soft  
inclusive momentum spectrum of inclusive $\Lambda_c$ production indicates that
$b\rightarrow c\bar cs$ is the dominant source of $\Lambda_c$'s in  
$B$ decays \cite{dcfw}. The hypothesis predicted (i) large wrong-sign $\ell^-  
\Lambda_c$ correlations, where the lepton comes from the semileptonic decay of  
one $B$ and the
$\Lambda_c$ from the other $B$ in an $\Upsilon (4S)$ event; and (ii)  
large $\Xi_c$ production in $B$ decays, which at that time had not been  
observed and was believed to be highly suppressed~\cite{crawford}.
Shortly afterwards, CLEO observed the first evidence of $\Xi_c$  
production in $B$ decays (see Table I), but also found that the right-sign  
$\ell^+\Lambda_c$ correlations are dominant (see Table III) 
\cite{glasgowbaryon}.
CLEO measured \cite{glasgowbaryon}
\begin{equation}
r_{\Lambda_c} \equiv \frac{B(\overline B\rightarrow  
\overline\Lambda_c
X)}{B(\overline B\rightarrow \Lambda_c X)} = 0.20 \pm 0.14 \;.
\end{equation}

Because the CLEO measurements of inclusive $\Xi_c$ production in $B$ decays
involve large uncertainties, and their central values appear to us to be too
high, this Appendix correlates $\Xi_c$ and $\Omega_c$ production  
in tagged $\stackrel{(-)}{B}$ decays to that of the more accurately measured
$\Lambda_c$. We neglect $b\rightarrow u$ transitions and use the Cabibbo
suppression factor $\theta^2 = (0.22)^2$ for charmed baryon production in
$b\rightarrow c\bar us (b\rightarrow c\bar cd)$ versus $b\rightarrow  
c\bar ud^\prime (b\rightarrow c\bar cs^\prime )$ transitions.
The parameter $p=0.15 \pm 0.05$ models the fraction of $s\bar s$ fragmentation
relative to $f\bar f$ fragmentation from the vacuum, where $f=u,d$ or $s$.
(This value for $p$ was chosen to demonstrate that even a
large value yields a 
significant reduction in $\Xi_c$ production in $\overline B$ decays.)

We denote by $C_{\bar ud}$ the fraction of $\overline B$'s which decay to 
weakly decaying charmed baryons which come from $b\rightarrow c\bar ud$,  
and define $C_{\bar cs}, C_{\bar cd}, C_{\bar us}$ analogously. Because our  
model allows for substantial charmless-baryon, charmless-anti-baryon production 
in $B$ decays, $C_{\bar ud}$ is smaller, or at most equal to $B_{\bar ud}$
as defined in  Ref.~\cite{dcfw}. Similar comments can be made for 
$C_{\bar cs}, C_{\bar cd}$, and $C_{\bar us}$ relative to $B_{\bar cs}, 
B_{\bar cd}$, and $B_{\bar us}$. The simplest version of the model predicts
\begin{equation} 
B(\overline B\rightarrow \Lambda_c X) = (1-p)(C_{\bar ud} + C_{\bar  
cd})
\end{equation}
\begin{equation}
B(\overline B\rightarrow \overline\Lambda_c X) = (1-p) (C_{\bar cs} +  
C_{\bar
cd})
\end{equation}
\begin{equation}
B(\overline B\rightarrow \Xi_c X) = p\; C_{\bar ud} + (1-p)\; C_{\bar  
us} + (1-p)\;
C_{\bar cs} + p\; C_{\bar cd}
\end{equation}
\begin{equation}
B(\overline B\rightarrow \overline\Xi_c X) = p\; (C_{\bar cs} +  
C_{\bar cd})
\end{equation}
\begin{equation}
B(\overline B\rightarrow \Omega_c X) = p(C_{\bar us} + C_{\bar cs})
\end{equation}
\begin{equation}
B(\overline B\rightarrow \overline\Omega_c X) = 0
\end{equation}

The Cabibbo structure

\begin{equation}
C_{\bar cd}/(C_{\bar cd} +C_{\bar cs})= C_{\bar cd}/C_{\bar  
cs^\prime} = \theta^2
\end{equation}
\begin{equation}
C_{\bar us}/(C_{\bar us}+C_{\bar ud}) = C_{\bar us}/C_{\bar  
ud^\prime} =
\theta^2
\end{equation}
allows us to express the six observables listed on the left-hand sides of
Eqs.~(5.2)-(5.7) in terms of the two unknowns $C_{\bar ud}$ and $C_{\bar cs}$.
The latter are in turn obtained from the two  
measurements involving inclusive $\Lambda_c$ production in $B$ decays,   
namely $Y_{\Lambda_c}$ and $r_{\Lambda_c}$, as follows~:

\begin{equation}
\frac{C_{\bar ud}}{Y_{\Lambda_c}}=\frac{(1+\lambda^2 -\lambda^2
r_{\Lambda_c})}{(1-p) (1+\lambda^2 )(1+r_{\Lambda_c})}\;,
\end{equation}
\begin{equation}
\frac{C_{\bar cs}}{C_{\bar ud}} = \frac{r_{\Lambda_c}}{1+\lambda^2
(1-r_{\Lambda_c})} \; ,
\end{equation}
where
\begin{equation}
\lambda^2 =\frac{\theta^2}{|V_{cs}|^2} =  
\frac{\theta^2}{\left(1-\frac{1}{2}
\theta^2 \right)^2} \; .
\end{equation}

The inclusive $\stackrel{(-)}{\Xi_c},
\stackrel{(-)}{\Omega_c}$ yields in $\overline B$ decays are thus  
correlated to inclusive
$\stackrel{(-)}{\Lambda_c}$ production,
\begin{equation}
\frac{B(\overline B\rightarrow \Lambda_c X)}{Y_{\Lambda_c}} =
\frac{1}{1+r_{\Lambda_c}} \;,
\end{equation}
\begin{equation}
\frac{B(\overline B\rightarrow \overline\Lambda_c
X)}{Y_{\Lambda_c}}=\frac{r_{\Lambda_c}}{1+r_{\Lambda_c}} \;,
\end{equation}
\begin{equation}
\frac{B(\overline B\rightarrow \Xi_c X)}{Y_{\Lambda_c}}=\frac{C_{\bar
ud}}{Y_{\Lambda_c}} \left\{p+(1-p)\lambda^2 +\frac{C_{\bar  
cs}}{C_{\bar ud}}
(1-p+p\lambda^2 )\right\}\;,
\end{equation}
\begin{equation}
\frac{B(\overline B\rightarrow \overline\Xi_c  
X)}{Y_{\Lambda_c}}=\frac{C_{\bar
ud}}{Y_{\Lambda_c}} \;\frac{C_{\bar cs}}{C_{\bar ud}} \;p(1+\lambda^2  
)\;,
\end{equation}
\begin{equation}
\frac{B(\overline B\rightarrow \Omega_c X)}{Y_{\Lambda_c}} = p  
\frac{C_{\bar
ud}}{Y_{\Lambda_c}} \left(\lambda^2 +\frac{C_{\bar cs}}{C_{\bar  
ud}}\right) \;,
\end{equation}
\begin{equation}
B(\overline B\rightarrow \overline\Omega_c X) = 0 \;.
\end{equation}

We have taken $p$ to be a universal quantity and have assumed that  
the initially produced charmed baryon retains its charm [and when  
applicable, strange] quantum number[s] through to its weakly  
decaying offspring. This is not justified but is conservative 
in that it yields an upper limit for $baryon_c$ production in $B$ decays.
We typically expect the initially produced charmed baryons (via  
$b\rightarrow c)$ to be highly excited, while this is not expected of their  
pair-produced antibaryons (via $b\rightarrow \bar u$ or $b\rightarrow \bar  
c)$~\cite{dalitz}.
That a sizable fraction of these highly excited charmed baryons could break up
into a charmed meson, a charmless baryon, and additional debris is irrelevant to
our discussion which focuses on weakly decaying charmed baryon
production in $B$ decays.
In contrast, it is important to note that $\Xi^r_c \rightarrow
\Lambda_c \overline K X$ could occur significantly [the superscript
$r$ denotes excited resonances].
This introduces an additional mechanism for $\Lambda_c$ production in
$\overline B$ decays, which may help explain the small measured value of
$r_{\Lambda_c}$. It also decreases the naive estimate for weakly
decaying $\Xi_c$ production. Because our predictions have not  
incorporated such
effects, they should be viewed strictly as upper limits for $\Xi_c $
production in $\overline B$ decays.

\begin{table}
\caption{Inclusive Charmed Hadron Production in $B$
Decays as Measured by CLEO}
\begin{tabular}{|c|c|c|}
$T$ & $Y_T \equiv B(\overline B\rightarrow TX) +
B(\overline B\rightarrow \overline T X)$ & Reference \\
\tableline
$D^0$ & $(0.645 \pm 0.025) \left[\frac{3.91\%}{B(D^0 \rightarrow K^-  
\pi^+)}\right]$ & \cite{hitoshi} \\
\tableline
$D^+$ & $(0.235 \pm 0.017) \left[\frac{9.3\%}{B(D^+ \rightarrow K^-  
\pi^+\pi^+
)}\right]$ & \cite{hitoshi} \\
\tableline
$D$ & $(0.883 \pm 0.038) \left[\frac{3.91\%}{B(D^0 \rightarrow  
K^-\pi^+)}\right]$ & \\
\tableline
$D_s$ & $(0.1211 \pm 0.0096) \left[\frac{3.5\%}{B(D_s \rightarrow  
\phi\pi )}\right]$ & \cite{menary} \\
\tableline
$\Lambda_c$ & $(0.030 \pm 0.005) \left[\frac{6\%}{B(\Lambda_c  
\rightarrow pK^-
\pi^+)}\right]$ & \cite{zoeller} \\
\tableline
$\Xi^+_c$ & $0.020 \pm 0.007$ & \cite{jain}\\
\tableline
$\Xi^0_c$ & $0.028 \pm 0.012$ & \cite{jain}

\end{tabular}
\end{table}

\begin{table}
\caption{Absolute Branching Ratios of Key Charm Decays}
\begin{tabular}{|c|c|c|}
Mode & BR [in \%] & Reference \\
\tableline
\tableline
$D^0 \rightarrow K^- \pi^+$ & $3.91 \pm 0.19$ & \cite{cleod0} \\
\tableline
$D_s\rightarrow \phi\pi$ & $3.5 \pm 0.4$ & \cite{pdg} \\
\tableline
$\Lambda_c \rightarrow pK^- \pi^+$ & $6.0 \pm 1.5$ & \cite{shipsey}
\end{tabular}
\end{table}

\begin{table}
\caption{Inclusive Charmed Hadron Production in Tagged $B$ Decays as  
Measured by
CLEO}
\begin{tabular}{|c|c|c|}
Observable & Value & Reference \\
\tableline
\tableline
$r_{\Lambda_c} \equiv \frac{B(\overline B\rightarrow  
\overline\Lambda_c X)}{B(\overline
B\rightarrow \Lambda_c X)}$ & $ 0.20 \pm 0.14$ & \cite{glasgowbaryon}  
\\
\tableline
$r_D \equiv \frac{B(\overline B\rightarrow \overline DX)}{B(\overline  
B\rightarrow
DX)}$ & $0.107 \pm 0.034 $ & \cite{moriond} \\
\tableline
$f_{D_s} \equiv \frac{B(\overline B\rightarrow D^+_s X)}{Y_{D_s}}$ &  
$0.172 \pm 0.083$ & \cite{cho}

\end{tabular}
\end{table}

\begin{table}
\caption{Extracted Values of $B(D^0\rightarrow K^-\pi^+)$.}
\label{table:dkpi}
\begin{tabular}{|l|l|}
\hline
\hline
Analysis			& $B(D^0\rightarrow K^-\pi^+)$ \\
\hline
$B(\overline{B}\rightarrow DX)$	& $(3.50\pm0.21)\%$ \\
$R_c(D^{*+})$			& $(3.35\pm0.25)\%$ \\
$R_c(\rm{charm~counting})$	& $(3.67\pm0.36)\%$ \\
Semileptonic BR's		& $(2.9\pm0.4)\%$ \\
\hline
ALL				& $(3.40\pm0.14)\%$ \\
\hline
\hline
\end{tabular}
\end{table}

\begin{table}
\caption{ $R_b$ Results as of the Brussels EPS-HEP-95 Conference }
\label{table:Rb}
\begin{tabular}{|l|l|l|l|}
\hline
\hline
Experiment & Data Set(s) & Measurement Type & $R_b$ ($R_c=0.172$ Fixed) \\
\hline
ALEPH  & 1992 & Lifetime & $0.2192\pm0.0022\pm0.0026$ \\
DELPHI & 1992-3 prel. & Lifetime & $0.2216\pm0.0017\pm0.0027$ \\
DELPHI & 1992-3 prel. & Mixed & $0.2231\pm0.0029\pm0.0035$ \\
DELPHI & 1992-3 prel. & Multivariate & $0.2186\pm0.0032\pm0.0022$ \\
OPAL   & 1992-4 prel. & Lifetime $+$ lepton & $0.2197\pm0.0014\pm0.0022$ \\
ALEPH  & 1990-1 & Event Shape & $0.228\pm0.005\pm0.005$ \\
SLD    &        & Lifetime & $0.2171\pm0.0040\pm0.0037$ \\
L3     & 1991   & Event Shape & $0.222\pm0.003\pm0.007$ \\
\hline
LEP    &        & Lepton Fits & $0.2219\pm0.0039$ \\
\bf{LEP+SLD}&        & \bf{Lifetime Fits} & $\bf{0.2200\pm0.0017}$ \\
\hline
\bf{ALL}    &        &  & $\bf{0.2205\pm0.0016}$ \\
\hline
\hline
\end{tabular}
\end{table}

\begin{table}
\caption{ Purity, efficiencies and correlation values.}
\label{table:btag}
\begin{tabular}{|l|l|l|l|l|l|}
\hline
\hline
Experiment & $B$ Purity & \eb & \ec & \euds & $C_b$ \\
\hline
ALEPH  & 0.96 & 0.26 & 1.18$\times10^{-2}$ & 0.88$\times10^{-3}$ & 0.943 \\
DELPHI & 0.92 & 0.21 & 1.60$\times10^{-2}$ & 2.52$\times10^{-3}$ & 0.952 \\
OPAL   & 0.94 & 0.23 & 1.37$\times10^{-2}$ & 1.01$\times10^{-3}$ & 1.006 \\
SLD    & 0.94 & 0.31 & 2.30$\times10^{-2}$ & 0.87$\times10^{-3}$ & 0.995 \\
\hline
\hline
\end{tabular}
\end{table}

\begin{references}
\bibitem{pvalue}
We choose here $p=0.10 \pm 0.05$ motivated by the measured primary
production fraction of $D_s$ mesons versus $D+D_s$ mesons. A larger 
value of $p$ was used in the Appendix to demonstrate conservatively that 
$Y_{\Xi_c}/Y_{\Lambda_c}$ is much smaller than currently believed.
\bibitem{dalitz}Consider the $b\rightarrow cd\bar u$, $V-A$ transition.
The invariant mass of the $cd$ peaks at the highest possible values as  
seen in a Dalitz plot \cite{bdy}. Thus we expect the charmed baryons
initially produced via $b\rightarrow c$ to be highly excited. In contrast, 
the $V-A$  nature of the interaction favors smaller energies for the 
$\bar u$ antiquark in the rest frame of the decaying
$b$. Since the spectator antiquark $\bar q$ of the  
$\overline B(\equiv b\bar q )$ meson involves only a modest Fermi momentum, 
the invariant mass of the $\bar u\bar q$ system is also expected to be modest.
\end{references}
\end{document}